\def \SAIT #1 #2 {{\em Mem.\ Soc.\ Astron.\ It.\/} {\bf #1}, #2}
\def \MESS #1 #2 {{\em The Messenger\/} {\bf #1}, #2}
\def \ASTRNACH #1 #2 {{\em Astron. Nach.\/} {\bf #1}, #2}
\def \AAP #1 #2 {{\em Astron. Astrophys.\/} {\bf #1}, #2}
\def \AAL #1 #2 {{\em Astron. Astrophys. Lett.\/} {\bf #1}, L#2}
\def \AAR #1 #2 {{\em Astron. Astrophys. Rev.\/} {\bf #1}, #2}
\def \AAS #1 #2 {{\em Astron. Astrophys. Suppl. Ser.\/} {\bf #1}, #2}
\def \AJ #1 #2 {{\em Astron. J.\/} {\bf #1}, #2}
\def \ANNREV #1 #2 {{\em Ann. Rev. Astron. Astrophys.\/} {\bf #1}, #2}
\def \APJ #1 #2 {{\em Astrophys. J.\/} {\bf #1}, #2}
\def \APJL #1 #2 {{\em Astrophys. J. Lett.\/} {\bf #1}, L#2}
\def \APJS #1 #2 {{\em Astrophys. J. Suppl.\/} {\bf #1}, #2}
\def \APSS #1 #2 {{\em Astrophys. Space Sci.\/} {\bf #1}, #2}
\def \ASR #1 #2 {{\em Adv. Space Res.\/} {\bf #1}, #2}
\def \BAIC #1 #2 {{\em Bull. Astron. Inst. Czechosl.\/} {\bf #1}, #2}
\def \JSQRT #1 #2 {{\em J. Quant. Spectrosc. Radiat. Transfer\/} {\bf #1}, #2}
\def \MN #1 #2 {{\em Mon. Not. R. Astr. Soc.\/} {\bf #1}, #2}
\def \MEM #1 #2 {{\em Mem. R. Astr. Soc.\/} {\bf #1}, #2}
\def \PLR #1 #2 {{\em Phys. Lett. Rev.\/} {\bf #1}, #2}
\def \PASJ #1 #2 {{\em Publ. Astron. Soc. Japan\/} {\bf #1}, #2}
\def \PASP #1 #2 {{\em Publ. Astr. Soc. Pacific\/} {\bf #1}, #2}
\def \NAT #1 #2 {{\em Nature\/} {\bf #1}, #2}

\documentstyle[twoside]{memsait}
\input epsf.sty
\begin{opening}
\title{THE FIRST HIGH-ENERGY X-RAY SPECTRUM OF A Z$>2$ RADIO-QUIET 
QUASAR: Q1101-264}
\author{C. VIGNALI$^1$, A. COMASTRI$^2$, M. CAPPI$^3$, G.G.C. PALUMBO$^{1,3}$}
\institute{$^1$Universit\`a di Bologna, via Zamboni 33, I-40126, Bologna,
Italy\\
$^2$Osservatorio Astronomico, via Zamboni 33, I-40126, Bologna, Italy\\
$^3$ITeSRE/CNR, via Gobetti 101, I-40129, Bologna, Italy}
\date{} 
\end{opening}

\begin{document}

\oddpagefooter{}{}{} 
\evenpagefooter{}{}{} 
\ 
\bigskip

\begin{abstract}
Results of the X-ray spectral analysis of the high-redshift radio-quiet quasar 
Q1101-264 are presented. The ASCA spectrum suggests a marginal evidence of a FeK$\alpha$ 
emission line at about 2 keV (observer's frame). 
Both the ASCA and ROSAT 
spectra are well fitted by a power law with spectral slope $\Gamma$$\sim$1.9. 
This is the first 0.3-30 keV spectrum (rest frame) of a z $>$ 2 
radio-quiet quasar.
\end{abstract}

\section{Data reduction} 
Q1101-264 is a high-redshift (z=2.15) radio-quiet quasar. It was 
observed with the gas imaging spectrometer (GIS) and solid
state spectrometer (SIS) on board the ASCA satellite (Tanaka et al. 1994)
in June 1996 and with the Rosat PSPC (Pfeffermann et al. 1987) in December 
1993. SIS grade 6 data were also included in order to improve 
the statistics above 5-6 keV (Mukai \& Weaver 1996). 
The GIS data had too low statistics for a spectral analysis and were, 
therefore, excluded from the following analysis. 
The total exposure times after screening were $\sim$ 17 Ks/SIS and 5 Ks for ROSAT PSPC. 

\section{Spectral analysis}
Flux variability of a factor of $\sim$ 2 on a time scale of about 
10 months (rest frame) is present from ROSAT and ASCA observations (tab. 1). 
The spectral analysis of ROSAT and ASCA data of Q1101-264 gives 
consistent results in terms of a power law 
continuum plus absorption, which is however poorly constrained because of 
the low statistics (tab. 1). The $N_{\rm H}$-$\Gamma$ confidence contours 
derived from the joint ROSAT + ASCA fit (fig. 1) show that an excess of 
absorption above the galactic value cannot be ruled out with the present data. 
Intrinsic absorption have been recently found in the X-ray spectra of 
high-redshift radio-loud quasar (Elvis et al. 1994, Serlemitsos et al. 1995, 
Cappi et al. 1997), but seems to be absent in high-redshift radio-quiet 
quasars (Elvis 1995). 
In this respect, Q1101-264 analysis cannot clarify the situation. 
The spectral slope derived from the joint ASCA and ROSAT fit 
($\Gamma$$\simeq$1.9$\pm{0.2}$) is consistent with that of lower z objects 
(Williams et al. 1992, Lawson \& Turner 1997). 
If Q1101-264 can be considered representative of the high-z 
radio-quiet objects, the present results suggest that the X-ray spectrum 
do not show any evolution with redshift. This favours a short-lived 
scenario, where quasars are active only for a short time (t$\sim$10$^{8}$ yr) 
accreting close to the Eddington limit. 
The most significant result, if confirmed, is the presence of the 
FeK$\alpha$ emission line, whose equivalent width results to be 
particularly high. A neutral or ionized reflection component 
is not required by, but consistent with, the data. 
Because of the presence of Si K and Au-M edges around 2 keV in the 
instrumental response, we checked whether the line could be due to 
remaining calibration uncertainties. 
The evidence of the line is robust also to reasonable energy scale shifts
which indicates that the feature might be real (tab. 1). 
\vspace{-0.1truecm}
\centerline{\bf Tab. 1 - Power law: spectral parameters}

\begin{table}[h]
\begin{tabular}{|l|c|c|c|c|c|c|c|}
\hline
\multicolumn{8}{|c|}{\bf Q1101-264}\\
\hline
detector&$N_{\rm H}$&$\Gamma$&$E_{{\rm K}\alpha}$&EW&$\chi^{2}/dof$&
$F_{0.4-2}$&$F_{2-10}$\\
&(10$^{20}$ cm$^{-2}$)&&(keV)&(eV)&&(10$^{-13}$)&(10$^{-13}$)\\
\hline
SIS0+1&$<$34.1&2.08$^{+0.56}_{-0.45}$&\dotfill&\dotfill&32.3/25&1.5&2.2\\
SIS0+1&5.68 fix.&1.90$^{+0.24}_{-0.23}$&\dotfill&\dotfill&32.8/26&1.6&2.4\\
SIS0+1&5.68 fix.&1.93$^{+0.27}_{-0.24}$&6.50$^{+0.18}_{-0.17}$
&728$^{+605}_{-577}$&30.2/24&1.6&2.4\\
\hline
PSPC&5.68 fix.&1.90$^{+0.34}_{-0.39}$&\dotfill&\dotfill&10.1/8&3.4&\dotfill\\
\hline
SIS0+1+PSPC&5.68 fix.&1.90$\pm{0.20}$&\dotfill&\dotfill&43/36&\dotfill
&\dotfill\\
\hline
\end{tabular}
\end{table}
%

%
\begin{figure}
\epsfysize=4.7cm 
\hspace{3.9cm}\epsfbox{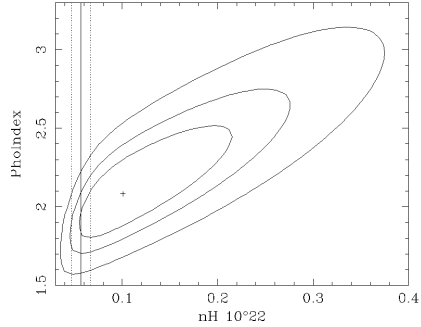}
\caption[h]{ASCA SIS0+1 + ROSAT $N_{\rm H}$-$\Gamma$ confidence contours.}
\end{figure}
%
%


\end{document}